\RequirePackage{fix-cm}

\documentclass[twocolumn,epjc3]{svjour3}          

\RequirePackage[T1]{fontenc}

\smartqed  

\RequirePackage{graphicx}
\RequirePackage{mathptmx}      
\RequirePackage{flushend}
\RequirePackage[colorlinks,citecolor=blue,urlcolor=blue,linkcolor=blue]{hyperref}

\RequirePackage[numbers,compress]{natbib}
\usepackage{amsmath}
\renewcommand{\mathcal}[1]{\text{\usefont{OMS}{cmsy}{m}{n}#1}}
\newcommand{\al}{\alpha}
\newcommand{\gm}{\gamma}
\newcommand{\Gm}{\Gamma}
\newcommand{\dl}{\delta}
\newcommand{\ep}{\varepsilon}
\newcommand{\lm}{\lambda}
\newcommand{\Uc}{\mathcal{U}}
\newcommand{\Fc}{\mathcal{F}}
\newcommand{\dd}{\mbox{d}}
\newcommand{\tmin}{\text{min}}
\newcommand{\tmax}{\text{max}}
\newcommand{\nn}{\nonumber}

\newcommand{\journalref}{Eur. Phys. J. C 72, 2139 (2012)}
\newcommand{\DOI}{10.1140/epjc/s10052-012-2139-2}
\newcommand{\arXiv}{1206.0546}
\newcommand{\preprints}{TTK-12-21, TTP12-015, SFB/CPP-12-31}
\def\makeheadbox{{%
\hbox to0pt{\vbox{\baselineskip=10dd\hrule\hbox
to\hsize{\vrule\kern3pt\vbox{\kern3pt
\hbox{Published in \href{http://dx.doi.org/\DOI}{\bfseries\journalref}}
\hbox{\href{http://arxiv.org/abs/\arXiv}{arXiv:\arXiv}, \preprints}
\kern3pt}\hfil\kern3pt\vrule}\hrule}%
\hss}}}

\journalname{Eur. Phys. J. C}

\begin{document}

\title{Expansion by regions:\\
   revealing potential and Glauber regions automatically}

\author{Bernd Jantzen\thanksref{e1,addr1}
        \and
        Alexander V. Smirnov\thanksref{e2,addr2,addr4}
        \and
        Vladimir A. Smirnov\thanksref{e3,addr3,addr4}
}

\thankstext{e1}{e-mail: jantzen@physik.rwth-aachen.de}
\thankstext{e2}{e-mail: asmirnov80@gmail.com}
\thankstext{e3}{e-mail: smirnov@theory.sinp.msu.ru}

\institute{Institut f\"ur Theoretische Teilchenphysik und Kosmologie,
    RWTH Aachen University, 52056 Aachen, Germany\label{addr1}
  \and
  Scientific Research Computing Center, Moscow State University,
    119992 Moscow, Russia\label{addr2}
  \and
  Skobeltsyn Institute of Nuclear Physics, Moscow State University,
    119992 Moscow, Russia\label{addr3}
  \and
  Institut f\"ur Theoretische Teilchenphysik, KIT,
    76128 Karlsruhe, Germany\label{addr4}
}

\date{Received: 9 June 2012 / Revised: 23 July 2012}

\maketitle

\begin{abstract}
  When performing asymptotic expansions using the strategy of expansion by
  regions, it is a non-trivial task to find the relevant regions. The
  recently published Mathematica code \texttt{asy.m} automates this task,
  but it has not been able to detect potential regions in threshold
  expansions or Glauber regions. In this work we present an algorithm and
  its implementation in the update \texttt{asy2.m} which also reveals
  potential and Glauber regions automatically.
\end{abstract}

\section{Introduction}

If a given Feynman integral depends on kinematic invariants and masses
which essentially differ in scale, a natural idea is to expand it
in ratios of small and large parameters. As a result, the integral is written
as a series of simpler quantities than the original integral itself and it can be
substituted by a sufficiently large number of terms of such an expansion.
For limits typical of Euclidean space (for example, the off-shell large-momentum limit
or the large-mass limit), one can write down the corresponding asymptotic expansion
in terms of a sum over certain subgraphs of a given
graph~\cite{Chetyrkin:1988zz,Chetyrkin:1988cu,Gorishnii:1989dd,
Smirnov:1990rz,Smirnov:1994tg,books1a}. This prescription of expansion by subgraphs
has been mathematically proven (see \cite{Smirnov:1990rz} and Appendix~B.2 of
\cite{books1a}).
Moreover, there is an automated tool~\cite{Seidensticker:1999bb,Harlander:1997zb}
where such an expansion by subgraphs is implemented.

For limits typical of Minkowski space (i.e.\ which cannot be formulated in
Euclidean space) the universal strategy of
expansion by regions~\cite{Beneke:1997zp,Smirnov:1998vk,Smirnov:1999bza,books1a}
is available. It consists of the following prescriptions:
\begin{itemize}
\item
  Divide the space of the loop momenta into various regions and, in
  every region, expand the integrand in a Taylor series with respect
  to the parameters that are considered small there.
\item
  Integrate the integrand, expanded in the appropriate way in every
  region, over the \emph{whole integration domain} of the loop momenta.
\item
  Set to zero any scaleless integral.
\end{itemize}
As shown in~\cite{Smirnov:1999bza}, this prescription can also be applied
to parametric representations of Feynman integrals, i.e.\ alpha parameters
(or generalized Feynman parameters) integrated from 0 to~$\infty$,
eventually restricted by a delta function. Then the regions are specified
by scaling relations between the parameters.

There is no mathematical proof that this prescription is correct in all
situations. But also no examples are known where a proper application of
the expansion by regions leads to wrong results. An indirect proof exists
for limits typical of Euclidean space because here the strategy of regions
is equivalent to the mathematically proven expansion by subgraphs.
A systematic study of the expansion by regions was presented recently
in~\cite{Jantzen:2011nz}. There it was shown explicitly and
illustrated using various one-loop examples that one can start from a
decomposition of a given integral into non-intersecting domains and arrive
at an expansion by regions in the above sense. This requires certain
conditions on the choice and completeness of the considered regions which
were derived in~\cite{Jantzen:2011nz}. As pointed out there, the appearance
of additional overlap contributions can be avoided by adequate choices of
the regions and regularization parameters.

While these findings provide some hints on the proper choice of the
regions, it remains a non-trivial task to actually reveal the typical
regions for a given limit. Usually, one starts from considering one-loop
examples, checks the results against known analytical results, then
proceeds in two loops etc.  One can also use the second version
\cite{FIESTA2} of the code {\tt FIESTA} \cite{FIESTA} to obtain numerically
several first terms of a given asymptotic expansion.

Recently an algorithm for an automatic search of regions was suggested and
implemented on a computer as the open source {\tt Mathematica} code
\texttt{asy.m}~\cite{asy}.
The algorithm uses a geometric approach based on finding the convex hull
of a set of points determined from a parametric representation of the
Feynman integral. In this way all possible sets of scalings for the
(Feynman) parameters are found which lead to non-vanishing (because
non-scaleless) integrals. These regions may then be used to expand the
parametric integral, or they can be translated into regions for expanding
the integral in loop-momentum space.

This code works successfully for a large variety of limits, at least in
cases where the function~$\Fc$ in the corresponding parametric
representation which depends on kinematic invariants and masses is only
composed of terms with the same sign. Moreover, it was shown in \cite{asy}
that in this case there are no regions except for the ones produced by the
code (all others result in scaleless integrals). In particular, the code
{\tt asy.m} works for Sudakov-type limits which are typical of Minkowski
space. As it was pointed out in \cite{asy} the code does not reveal
potential regions in threshold expansions, and similarly it fails to detect
the so-called Glauber regions.

Our goal in the present paper is to provide an algorithm and the
corresponding update {\tt asy2.m} of the code \texttt{asy.m} which
automatically identifies all regions relevant for a given integral,
including potential and Glauber regions.
We start in Section~\ref{sec:preliminaries} by introducing the parametric
representations of loop integrals which we use later and by explaining how
the contribution of a given region is obtained in the language of such
parametric integrals.
Then we elaborate our algorithm for revealing potential and Glauber regions
and explain how the code \texttt{asy2.m} is applied in such cases. This is
done in Section~\ref{sec:potential} for an example with a potential region,
and Section~\ref{sec:Glauber} deals with the more complicated problem of
revealing Glauber regions.
In both Sections \ref{sec:potential} and~\ref{sec:Glauber}, we first
formulate simple changes of variables and decompositions of a given Feynman
integral, using instructive one-loop examples, which lead to integrals
where {\tt asy2.m} is able to detect the relevant regions and print their
list in terms of the scalings of the parameters. Then, for both cases, we
explain how to use the new features of {\tt asy2.m} to perform these
algorithmic steps automatically.

In the case of Section~\ref{sec:Glauber} with Glauber regions, the
structure of the regions differs depending on whether the expansion is
performed in loop-momentum space or at the level of the parametric
integral. We show in Section~\ref{sec:Glauber_l} how to disentangle and
match the various regions arising in this problem by using generic
propagator powers, and how {\tt asy2.m} can be employed to automate such an
analysis.

A summary of the new features and the syntax of \texttt{asy2.m} (together
with a download link) is provided in Section~\ref{sec:syntax}. In
Section~\ref{sec:conclusion} we conclude by discussing the mathematical
problem of proving the expansion by regions for a simple example which is
not related to Feynman integrals, but where \texttt{asy2.m} works
successfully.

\section{Expansion by regions in parametric representations}
\label{sec:preliminaries}

We are dealing with dimensionally regularized Feynman integrals
\begin{align}
  F(q_1,\ldots,q_n;a_1,\ldots,a_N;d) &=
  \idotsint \prod_{i=1}^h \dd^d k_i \, \frac{1}{\prod_{l=1}^N E_l^{a_l}}\,,
  \label{eqbn-d1}
\end{align}
where $h$ is the number of loops, the indices $a_l$ are general powers of
the propagators, the dimension is $d=4-2\ep$ and the denominators $E_l$ are
given by
\begin{eqnarray}
 E_{l}&=&\sum_{i\geq j \geq 1}^h A^{i j}_{l} \, k_i \cdot k_j
+ \sum_{i=1}^h B^{i}_{l}\cdot  k_i
+ D_l
+ i0 \;,
\label{denom-d1}
\end{eqnarray}
i.e.\ they are quadratic or linear functions of the external momenta $q_i$
and the loop momenta $k_i$ with the usual infinitesimal imaginary part
$+i0$. Monomials in the numerator are taken into account as denominators
raised to negative powers.

The alpha representation of~(\ref{eqbn-d1}) takes the form
\begin{align}
&F(q_1,\ldots,q_n;a_1,\ldots,a_N;d) =
(i \pi^{d/2})^h \,
\frac{e^{-i \pi (a + h d/2)/2}}{\prod_{l=1}^N \Gm(a_l)}
\nn \\ &\qquad \times
\int_0^\infty \cdots \int_0^\infty
\prod_{l=1}^N \left(\dd\al_l \, \al_l^{a_l-1}\right)
\Uc^{\;-d/2} \, e^{-i \,\Fc/\Uc}
\; ,
\label{alpha-d}
\end{align}
where $a=\sum_l a_l$.
The functions $\Uc$ and $\Fc$ depend polynomially on the alpha
parameters~$\al_l$. Furthermore, $\Uc$ and $\Fc$ are homogeneous functions
of the alpha parameters with the homogeneity degrees $h$ and $h+1$,
respectively. The function~$\Fc$ is linear in the kinematic invariants
and/or squared masses which we denote by $s_i,\;i=1,2,\ldots$, while the
function~$\Uc$ is independent of the $s_i$.

If (\ref{eqbn-d1}) is an integral with standard propagators
$1/(p_l^2-m_l^2+i0)$ associated with the lines of a graph, then the
functions $\Uc$ and $\Fc$ are called Symanzik polynomials and are given by
the well-known formulae in terms of trees and $2$-trees.
For a general Feynman integral of the form~(\ref{eqbn-d1}) one can obtain
these functions using the simple public code \texttt{UF.m}\footnote{%
  The function \texttt{UF[]} from \texttt{UF.m}~\cite{UF.m} is called with
  three arguments: The list of loop momenta, the list of denominators of
  the propagators and a list of replacement rules for all kinematic
  invariants. The output is a list with the following entries: the
  function~$\Uc$, the function~$\Fc$ and the number of loops.
  In order to obtain $\Uc$ and $\Fc$ with the correct sign, denominators
  have to be specified with the opposite sign as in~(\ref{denom-d1}), i.e.\
  corresponding to a negative imaginary part $-i0$:
  \texttt{UF[\{k1,k2,...\}, \{-E1,-E2,...\},
    \{\textrm{replacement rules}\}]}.%
}~\cite{UF.m} which is also part of the codes \texttt{asy.m} and \texttt{asy2.m}.

If some of the indices~$a_l$ are negative integers, i.e.\ they correspond to
numerators instead of denominators of the integral~(\ref{eqbn-d1}), the
alpha representation~(\ref{alpha-d}) is to be understood in the limit where
these indices tend to their negative integer values. Effectively, the
integration over the corresponding parameters~$\al_l$ is replaced by
differentiating with respect to these parameters and setting them to zero.

Closely related to (\ref{alpha-d}) is the (generalized) Feynman parametric
representation
\begin{align} 
&F(q_1,\ldots,q_n;a_1,\ldots,a_N;d) =
(i \pi^{d/2})^h \,
\frac{e^{-i \pi a} \, \Gm(a - h d/2)}{\prod_{l=1}^N \Gm(a_l)}
\nn \\ &\times
\int_0^\infty \dd x_1 \cdots \int_0^\infty\dd x_N \,
\delta\!\left( \sum_{l\in \nu} x_l-1\right)
I(x_1,\ldots,x_N;s_1,s_2,\ldots)
\; ,
\label{alpha-d-mod}
\end{align}
where $\nu$ in the argument of the delta function is an arbitrary non-empty
subset of $\{1,\ldots,N\}$,
\begin{equation}
  I(x_1,\ldots,x_N;s_1,s_2,\ldots)= \prod_{l=1}^N x_l^{a_l-1} \,
  \Uc^{\;a-(h+1) \frac{d}{2}} \, (\Fc - i0)^{h \frac{d}{2}-a}
\label{integrand}
\end{equation}
and the functions $\Uc$ and $\Fc$ are the same as those in~(\ref{alpha-d})
with the parameters $\al_l$ replaced by $x_l$.
It is well known that the formula~(\ref{alpha-d-mod}) holds for any choice
of the subset~$\nu$ in the argument of the delta function.\footnote{%
  See e.g.\ the discussion in Section~3.4 of~\cite{Smirnov:2006ry}.%
} This feature is related to the above-mentioned homogeneity properties of
the functions $\Uc$ and $\Fc$.\footnote{%
  See also (\ref{delta_int}), (\ref{delta_int_hom}) and
  Footnote~\ref{fn:delta_scaling} (p.~\pageref{fn:delta_scaling}) for a
  general proof.%
} If one chooses $\nu = \{1,\ldots,N\}$, the standard Feynman
parametrization is recovered.

Let us suppose that we have to study the asymptotic behaviour in a
one-scale limit, i.e.\ every mass and kinematic invariant has a certain
scaling $s_i\to s'_i= \rho^{\kappa_i} s_i$, $i=1,2,\ldots$, expressed in
powers of the small parameter of the problem, $\rho$.
The strategy of expansion by regions formulated in terms of parametric
integrals (\ref{alpha-d}) or~(\ref{alpha-d-mod})
\cite{Smirnov:1999bza,books1a} states that the asymptotic expansion in such
a limit is given by a sum over regions which are specified by the scalings
of the parameters $\al_l$ or $x_l$ expressed in powers~$r_l$ of the
expansion parameter~$\rho$. Each region~$r$ is labelled by the list
$r=\{r_1,\ldots,r_N\}$ of its scaling powers.
The contribution of the region~$r$ is obtained by scaling the masses and
kinematic invariants according to the given limit as specified above, by
substituting $\al_l \to \al'_l= \rho^{r_l} \al_l,\;l=1\ldots,N$, in the
integrand of~(\ref{alpha-d}) or $x_l \to x'_l= \rho^{r_l} x_l$ in the
integrand of~(\ref{alpha-d-mod}) and by expanding the integrand in powers
of~$\rho$. Here the product of the differentials $\dd\al_l$ or $\dd x_l$
provides another factor $\rho^{\sum_l r_l}$ to the power counting.

Explicitly, the contribution of the region~$r$ is given by the prefactor in
(\ref{alpha-d-mod}) times $\rho^{\sum_l r_l}$ times the integral
\begin{equation}
\int_0^\infty \dd x_1 \cdots\int_0^\infty\dd x_N \,
\delta\!\left( \sum_{l\in \nu} x'_l-1\right)
I(x'_1,\ldots,x'_N;s'_1,s'_2,\ldots)
\label{alpha-d-mod-scaled}
\end{equation}
with the integrand expanded in powers of~$\rho$. This expansion also
involves the argument of the delta function in~(\ref{alpha-d-mod-scaled}),
such that, under the expansion, certain parameters drop out of the argument
of the delta function and are integrated from 0 to~$\infty$. For this
reason the upper integration limit of all Feynman parameters should be kept
at infinity and not switched to~$1$ even if, before the expansion, their
integration is restricted by the delta function. One may avoid expanding the delta
function by choosing the original subset~$\nu$ in~(\ref{alpha-d-mod})
sufficiently small.

Let us write down the leading-order (LO) contribution of a given region in
a more explicit way. For the two basic functions in~(\ref{integrand}) we
have
\begin{align}
\Uc(x'_1,\ldots,x'_N) &=
  \sum_{j=u_\tmin}^{u_\tmax} \rho^j \, \Uc_j(x_1,\ldots,x_N)\;,
\nn \\
\Fc(x'_1,\ldots,x'_N;s'_1,s'_2,\ldots) &=
  \sum_{j=f_\tmin}^{f_\tmax} \rho^j \, \Fc_j(x_1,\ldots,x_N;s_1,s_2,\ldots)\;,
\label{lpolyn}
\end{align}
where the arguments of the polynomials $\Uc$ and $\Fc$ on the left-hand
side indicate that they are expressed in terms of the scaled parameters
$x'_l$ and $s'_i$, while their expansion coefficients on the right-hand
side are expressed in terms of $x_l$ and $s_i$.
According to the prescription formulated above, the LO contribution of the
region~$r$ is represented as
\begin{align}
&\rho^{\sum_l r_l a_l + u_\tmin \left(a-(h+1)\frac{d}{2}\right)
  + f_\tmin \left(h \frac{d}{2} - a\right)}
\nn \\ &\times
(i \pi^{d/2})^h \,
\frac{e^{-i \pi a} \, \Gm(a - h d/2)}{\prod_{l=1}^N \Gm(a_l)}
\int_0^\infty \cdots \int_0^\infty
\prod_{l=1}^N \left(\dd x_l \, x_l^{a_l-1}\right)
\nn \\ &\times
\delta\!\left( \sum_{l\in \nu_0} x_l-1\right)
\Uc_{u_\tmin}^{\;a-(h+1)\frac{d}{2}} \,
(\Fc_{f_\tmin}-i0)^{h \frac{d}{2} - a}
\;.
\label{alpha-d-mod0}
\end{align}
In principle, the argument of the delta function in~(\ref{alpha-d-mod0})
contains a sum over only those scaled parameters $x'_l = \rho^{r_l} x_l$ with
the minimal scaling power $r_l = r_\tmin = \min\{r_1,\ldots,r_N\}$. But by
rescaling $x_l \to \rho^{-r_\tmin} \, x_l \, \forall l$, the delta function
is transformed into its standard form without powers of~$\rho$, while the
rest of the integral remains invariant due to the homogeneity of the
polynomials $\Uc_{u_\tmin}$ and $\Fc_{f_\tmin}$. Finally, as for the
original Feynman parametric representation~(\ref{alpha-d-mod}), one can
choose again an arbitrary non-empty subset~$\nu_0$ for the sum in the
argument of the delta function in~(\ref{alpha-d-mod0}).

The list of scalings $r=\{r_1,\ldots,r_N\}$ of a region is only determined
up to adding the same arbitrary real number~$c$ to each entry, because the
corresponding contribution stays the same under $r_l \to r_l + c \, \forall
l$. In particular, the LO behaviour presented in~(\ref{alpha-d-mod0}) is
independent of such a shift~$c$, because $u_\tmin \to u_\tmin + hc$ and
$f_\tmin \to f_\tmin + (h+1)c$, due to the homogeneity properties of $\Uc$
and~$\Fc$.

The terms of the expansion come from various regions and can be ordered
according to accompanying powers of~$\rho$. After keeping some first terms
of the expansion one can set $\rho=1$ and write down the given Feynman
integral as these selected first terms plus a remainder which vanishes
sufficiently fast in the given limit.

It turns out that only a limited number of regions contribute to the
expansion because for the rest of the regions one obtains integrals without
scale which are set to zero. It is the subject of this paper and the task
of the code {\tt asy.m} and its updated version {\tt asy2.m} to find all
relevant regions for a given integral.

\section{Revealing potential contributions}
\label{sec:potential}

Let us consider the one-loop propagator diagram with two massive lines in
the threshold limit, i.e.\ when $y=m^2-q^2/4\to 0$ with $q$ being the external
momentum:
\begin{equation}
  F(q^2,m^2)=\int \frac{\dd^d k}{(k^2-m^2)\; \bigl((k-q)^2-m^2\bigr)}\;,
  \label{prop1_integral}
\end{equation}
where the usual $+i 0$ is implied in all the propagators.
Within the strategy of expansion by regions, the hard and the potential
regions give contributions to the expansion \cite{Beneke:1997zp,books1a}.
The previous version of the code {\tt asy.m} \cite{asy} reported only about
the hard region. The reason for this can be seen in the corresponding
parametric representation,
\begin{align}
  F(q^2,y)&=i\pi^{d/2} \, \Gm(\ep)
\nn \\* &\times
  \iint \frac{(x_1+x_2)^{2\ep-2}\;\delta\left(x_1+x_2-1\right)\;
  \dd x_1\dd x_2
  }{\left[
  \frac{q^2}{4}(x_1-x_2)^2 + y(x_1+x_2)^2
  -i 0\right]^{\ep}}\;,
  \label{prop1_alpha}
\end{align}
where the parameters~$x_i$ are integrated from 0 to $\infty$ (restricted by
the delta function).
As it was pointed out in \cite{asy}, it is the region where
$x_1\approx x_2$ (more precisely \mbox{$x_1-x_2 \sim y^{1/2}$}) which causes problems.
In other words, the polynomial in the square brackets
in~(\ref{prop1_alpha}) (considered at positive $q^2$ and $y$) has terms of
different sign, such that cancellations occur because of the presence
of the negative term $-q^2 x_1x_2/2$.

To reveal the missing potential contribution, let us perform a simple trick.
We decompose the integration domain into two subdomains, $x_1\leq x_2$ and
$x_2\leq x_1$. The two resulting integrals are equal to each other, but
such an equality will not generally take place for any integral.
In the first domain we turn to new variables by
$x_1=x_1'/2,\;x_2=x'_2+x'_1/2$, remove the primes at $x_i$ and obtain
the integral (again from 0 to $\infty$ with the usual restrictions
via the delta function)
\begin{equation}
  i\pi^{d/2} \, \frac{\Gm(\ep)}{2}
  \iint \frac{(x_1+x_2)^{2\ep-2}\;\delta\left(x_1+x_2-1\right)\;
  \dd x_1\dd x_2}
  {\left[
  \frac{q^2}{4}x_2^2 + y(x_1+x_2)^2
  -i 0\right]^{\ep}}\;.
  \label{prop1_alpha_1}
\end{equation}
The goal of this trick was to make the line $x_1=x_2$ (in the old variables)
the border of an integration domain which turned out to be
(in the new variables) $x_2=0$. Now we can run the code {\tt asy2.m}.
Since this is a parametrical integral rather than a Feynman integral
we use the newly introduced command\footnote{The name of the command refers
  to its application to parametric integrals contributing to Wilson loops.}
{\tt WilsonExpand[]} for integrals where all parameters are integrated from
0 to $\infty$:
\begin{verbatim}
WilsonExpand[q^2/4*x2^2 + y*(x1 + x2)^2,
  x1 + x2, {x1, x2}, {q -> 1, y -> x},
  Delta -> True]
\end{verbatim}
The first two arguments of \texttt{WilsonExpand[]} are the polynomials
$\Fc$ and $\Uc$, respectively, as defined in
Section~\ref{sec:preliminaries}. They can easily be determined from the
square brackets in the denominator of the parametric
integral~(\ref{prop1_alpha_1}) and from the round brackets in the
numerator.
The third argument is the list of integration parameters, and the fourth
argument specifies the scaling of the kinematic quantities with respect to
the small parameter which is labelled by the global symbol~\texttt{x}. Here
by \verb@y -> x@ we tell the code that $y$ is the small expansion
parameter, and by \verb@q -> 1@ we specify that the momentum~$q$ scales as
$y^0=1$.
The option \verb@Delta -> True@ tells \texttt{WilsonExpand[]} that, under
the integration, the sum over an arbitrary non-empty subset of the
integration parameters is restricted to~1 by a delta function.

Note that \texttt{WilsonExpand[]} can only take into account such a delta
function if the specific choice of the sum over parameters in the argument
of the delta function is irrelevant. This is the case for the generalized
Feynman parametric integral~(\ref{alpha-d-mod}) introduced in
Section~\ref{sec:preliminaries}. The integrals (\ref{prop1_alpha})
and~(\ref{prop1_alpha_1}) are special cases of~(\ref{alpha-d-mod}) such
that we could e.g.\ replace $\delta(x_1+x_2-1)$ by $\delta(x_1-1)$ without
changing the value of the integrals. If, however, a specific form of the
delta function is assumed, e.g.\ by replacing $x_1+x_2 \to 1$ under the
integral, then the option \texttt{Delta} of \texttt{WilsonExpand[]} does
not apply (see Section~\ref{sec:syntax} for details). Alternatively,
\texttt{WilsonExpand[]} can be used without the option \texttt{Delta} after
eliminating one of the integrations, e.g.\ via $\delta(x_1-1)$.

The output of the above-mentioned call of \texttt{WilsonExpand[]} is a list
of regions specified by the scaling of the parameters~$x_i$ in powers of
the small parameter~$y$:
\begin{verbatim}
{{0, 0}, {0, 1/2}}
\end{verbatim}
This specification of a region corresponds to the list of scalings
$\{r_1,\ldots,r_N\}$ introduced in Section~\ref{sec:preliminaries}.
The first entry in the output, $\{0,0\}$, refers to the hard region with
the scaling $x_1\sim y^0$, $x_2\sim y^0$ which has already been found by
\texttt{asy.m}. But now also the region $\{0, 1/2\}$ is found with the
scaling $x_1\sim y^0$, $x_2\sim y^{1/2}$ which provides the potential
contribution.

The contribution of the hard region starts with the order $y^0$. Every term
of the expansion can be evaluated in terms of gamma functions for
general~$\ep$.
According to the prescriptions for writing down the contribution of a
region formulated in Section~\ref{sec:preliminaries}, the contribution of
the $k$-th order expansion of~(\ref{prop1_alpha_1}) in the potential region
reads
\begin{equation}
  i\pi^{d/2} \, \frac{\Gm(\ep)}{2 \, k!}
  \iint \dd x_1\dd x_2 \, x_2^k
  \left(\frac{\partial}{\partial x_1}\right)^k \,
  \frac{x_1^{2\ep-2}\;\delta\left(x_1-1\right)}
    {\left(\frac{q^2}{4}x_2^2 + y \, x_1^2\right)^{\ep}}\;.
  \label{prop1_alpha_p}
\end{equation}
Only the leading order ($k=0$),
\begin{equation}
  i\pi^{d/2} \, \frac{\Gm(\ep)}{2}
  \int_0^\infty
  \frac{\dd x_2}{\left(\frac{q^2}{4}x_2^2 + y\right)^{\ep}}\;,
  \label{prop1_alpha_p0}
\end{equation}
yields a non-vanishing contribution which is evaluated
in terms of gamma functions at general $\ep$. Taking into account that
we have two identical integrals after our decomposition, we arrive at
the following result for the potential contribution which is of order $y^{1/2-\ep}$:
\begin{equation}
  i \pi^{d/2} \, \Gm(\ep-1/2)
  \sqrt{\frac{\pi y}{q^2}} \, y^{-\ep}
  \label{prop1_pot}
\end{equation}
in agreement with~\cite{books1a}.

In fact, such a trick of making manifest squares of some linear combination
of the integration parameters was already used in the code
{\tt FIESTA} \cite{FIESTA} in order to evaluate numerically Feynman integrals
at a threshold. Using the implementation of this procedure in {\tt FIESTA}
it turned out to be possible to automate the above trick for a general Feynman
integral.
In the present version {\tt asy2.m}, the user may call the command
\texttt{AlphaRepExpand[]} with the additional option \texttt{PreResolve}
enabled which automatically looks for the change of variables described above:
\begin{verbatim}
AlphaRepExpand[{k},
  {k^2 - m^2, (k - q)^2 - m^2},
  {q^2 -> qq, m^2 -> qq/4 + y},
  {qq -> 1, y -> x},
  PreResolve -> True]
\end{verbatim}
As in the previous version of \texttt{asy.m}, the arguments of
\texttt{AlphaRepExpand[]} are the list of loop momenta, the list of
denominators of the loop integral, a list of replacements for the kinematic
quantities, and the list of scalings with respect to the small
parameter~\texttt{x}.
The output is a list of entries which indicate for each region the changes of
variables, the Jacobian of the integral transformation, and the scalings of
the new variables:
\begin{verbatim}
{{{x[1] -> y[1] + y[2]/2, x[2] -> y[2]/2},
   2, {0, -1/2}},
 {{x[1] -> y[1] + y[2]/2, x[2] -> y[2]/2},
   2, {0, 0}},
 {{x[1] -> y[1]/2, x[2] -> y[1]/2 + y[2]},
   2, {0, 0}},
 {{x[1] -> y[1]/2, x[2] -> y[1]/2 + y[2]},
   2, {0, 1/2}}}
\end{verbatim}
The original Feynman parameters are labelled \verb@x[1]@, \verb@x[2]@,
$\ldots$, while the new parameters are labelled \verb@y[1]@, \verb@y[2]@,
$\ldots$. The scaling relations which determine the regions are specified
for the new parameters.

As explained in Section~\ref{sec:preliminaries}, a region remains invariant
if all of its scalings are shifted by the same amount, i.e.\ if all
parameters are rescaled simultaneously. So the scaling $\{0,-1/2\}$ shown
above is equivalent to $\{1/2,0\}$, i.e.\ the first new parameter is
suppressed by $y^{1/2}$ with respect to the second one.
For both parts of the integral decomposition the hard and potential regions
are found.

The preresolution algorithm implemented in {\tt asy2.m} (and switched on
with the option \texttt{PreResolve}) tries to eliminate factorized
combinations of terms in the function $\Fc$ which potentially
cancel each other, like $(x_1-x_2)^2$ in the example above. It checks all
pairs of variables (say, $x$ and $y$) which are part of monomials with
opposite sign. For all those pairs the code tries to build a linear
combination $z$ of $x$ and $y$ such that in the variables $x$ and $z$ or
$y$ and $z$ this monomial disappears. The code checks whether in the new
variables the number of monomials with opposite sign decreases. For all
such pairs the code recursively repeats the initial procedure in the new
variables. As a result it creates a tree of possible bisections and
corresponding replacements of variables. A leaf of this tree is a set of
sectors and functions such that one cannot decrease the number of monomials
with opposite sign any longer. Ideally it means that all monomials now have
the same sign. The code analyzes all leafs and chooses one of those with
the minimal number of opposite-sign monomials (or the minimal number of
sectors if the numbers of monomials with opposite sign coincide).
After finishing with the preresolution, the code performs the replacements
and looks for regions in all those sectors, using the algorithm of the
original code \texttt{asy.m} described in~\cite{asy}.

Note that the algorithm can only find the necessary variable
transformations if it is able to determine the relative signs of all terms
in the polynomial~$\Fc$. As the signs of symbols are unknown to
\texttt{asy2.m}, the substitution rules of the third and fourth arguments
of \texttt{AlphaRepExpand[]} must replace all kinematic quantities by
numbers (integers or fractions of integers) or powers of the small
parameter~\texttt{x}.

We have checked that the updated version {\tt asy2.m} works in various
examples of the threshold expansion (considered
in~\cite{Beneke:1997zp,books1a}): a triangle, a box, the two-loop
propagator diagram (with the masses $m,m,m,m,0$), a two-loop vertex
diagram. Because of the decomposition of a given integration domain into
subdomains, the number of resulting integrals for various regions increases
a little bit. For example, the (hard-hard) region for the two-loop
propagator diagram is described by six integrals, the (potential-ultrasoft)
region is also described by six integrals, etc. However, the
(potential-hard) region is described by four integrals with some regions
(with scalings composed of powers 1, 1/2 and 0), and four more integrals
with a set of regions of a different type (composed of 1 and 0).

Let us finally mention that the preresolution algorithm also works for
threshold expansions with unequal masses. Returning to the one-loop
example~(\ref{prop1_integral}), but now with different masses $m_1$ and
$m_2$ in the propagators and the expansion parameter defined as $y =
(m_1+m_2)^2/4 - q^2/4 \to 0$, we call:
\begin{verbatim}
AlphaRepExpand[{k},
  {k^2 - m1^2, (k - q)^2 - m2^2},
  {q^2 -> (m1 + m2)^2 - 4*y},
  {y -> x, m1 -> 2, m2 -> 3/2},
  PreResolve -> True]
\end{verbatim}
We have set the values of the masses to different rational numbers, $m_1
\to 2$ and $m_2 \to 3/2$ (the actual values are irrelevant), thus
permitting the preresolution algorithm to know about their sign without
assuming any equality or other relation between them. The
polynomial~$\Fc$ in the parametric representation of the Feynman
integral reads $\Fc = (m_1 x_1 - m_2 x_2)^2 + 4y x_1 x_2$, and
cancellations occur where $m_1 x_1 = m_2 x_2$. These cancellations are
automatically made explicit by adequate changes of variables. The output
\begin{verbatim}
{{{x[1] -> y[1] + 3*y[2]/7,
   x[2] -> 4*y[2]/7}, 7/4, {0, -1/2}},
 {{x[1] -> y[1] + 3*y[2]/7,
   x[2] -> 4*y[2]/7}, 7/4, {0, 0}},
 {{x[1] -> 3*y[1]/7,
   x[2] -> 4*y[1]/7 + y[2]}, 7/3, {0, 0}},
 {{x[1] -> 3*y[1]/7,
   x[2] -> 4*y[1]/7 + y[2]}, 7/3, {0, 1/2}}}
\end{verbatim}
shows that \texttt{asy2.m} detects both regions. The variable
transformations are always normalized such that sums of the parameters
(here $x_1+x_2$) remain invariant.

So, we now have a manifestly Lorentz-invariant treatment of threshold
expansion and a code that automatically provides the set of relevant regions.

\section{Revealing Glauber contributions}
\label{sec:Glauber}

\begin{figure}
  \centering
  \includegraphics[width=\columnwidth]{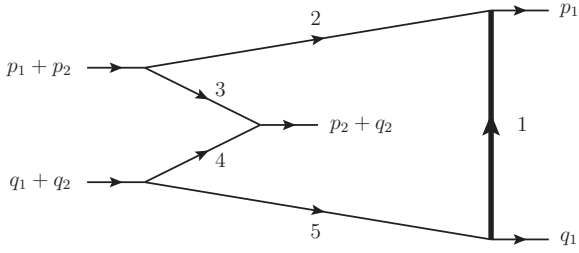}
  \caption{One-loop five-point integral exhibiting a Glauber contribution}
  \label{fig:Glauber}
\end{figure}
Let us consider the one-loop five-point integral in Fig.~\ref{fig:Glauber},
where two initial-state partons both perform a collinear splitting into two
partons each with momenta $p_1,p_2$ and $q_1,q_2$, respectively. While two
partons, one of each pair, collide with a large centre-of-mass energy
$Q=\sqrt{(p_2+q_2)^2}$, the two remaining partons exchange a particle with
the small mass~$m$.
We will use the simplified kinematics
$p_1 = p_2 = p$ and $q_1 = q_2 = q$ with $p^2 = q^2 = 0$
and $(p+q)^2 = 2p\cdot q = Q^2$ in the limit  $m^2/Q^2 \to 0$:
\begin{align}
  F(Q^2,m^2)&=\int \frac{\dd^d k}{(k^2-m^2)(k^2-2p\cdot k) (k^2+2p\cdot k)}
\nn \\* &\times
  \frac{1}{(k^2-2q\cdot k) (k^2+2q \cdot k)}
  \;.
  \label{5pt_integral}
\end{align}

Before we search for regions using \texttt{asy.m}, we notice that this
five-point integral is similar to the Sudakov form factor example
treated in Section~6 of~\cite{Jantzen:2011nz}. From the viewpoint of the
convergence of the expansions, the second and third propagators
of~(\ref{5pt_integral}) are equivalent, and so are the fourth and fifth
propagators. Effectively, the five-point integral has only three different
types of propagators which are equivalent to the ones of the three-point
integral in~\cite{Jantzen:2011nz}. So the integral~(\ref{5pt_integral}) can
be expanded in loop-momentum space employing the regions and convergence
domains known from~\cite{Jantzen:2011nz} (and using generic propagator
powers as analytic regulators where necessary):
\begin{itemize}
\item a hard region where $k \sim Q$,
\item a 1-collinear region where $k^2 \sim p\cdot k \sim m^2$ and $q\cdot k
  \sim Q^2$,
\item a 2-collinear region where $k^2 \sim q\cdot k \sim m^2$ and $p\cdot k
  \sim Q^2$,
\item a Glauber region where $p\cdot k \sim q\cdot k \sim m^2$, and the
  components of~$k$ perpendicular to the plane spanned by $p,q$ scale as
  $k_\perp \sim m$.
\end{itemize}
The collinear-plane region mentioned in~\cite{Jantzen:2011nz} yields
only scaleless contributions. But, in contrast to the three-point integral,
the five-point integral has a non-vanishing Glauber contribution. The
Glauber region even provides the leading contribution scaling as
$(m^2)^{-2-\ep}$, whereas the collinear contributions start with
$(m^2)^{-1-\ep}$ and the hard contribution starts with $(m^2)^0$.

The five-point integral~(\ref{5pt_integral}) can be represented in terms of
an integral over Feynman parameters,
\begin{align}
  &F(Q^2,m^2)=-i\pi^{d/2} \, \Gm(3+\ep)
  \idotsint \dd x_1\cdots\dd x_5
\nn \\* &\;\;\times
  \frac{\delta\left(\sum_i x_i-1\right) \; (x_1+\ldots+x_5)^{1+2\ep}
  }{\bigl[x_1(x_1+\ldots+x_5) m^2
  +(x_2-x_3)(x_4-x_5)Q^2-i 0\bigr]^{3+\ep}
   }\;,
  \label{Fp_integral}
\end{align}
where one can choose the sum in the argument of the delta function in an
appropriate way, i.e.\ restrict only the sum over a subset of the
parameters to 1 and extend the integration over the rest of the parameters
to the whole domain $[0,\infty)$.

Applying the strategy of expansion by regions in Feynman-parameter space
and trying to reveal regions relevant to the given limit $m^2/Q^2 \to 0$
with the help of the code {\tt asy.m} \cite{asy}, we call:
\begin{verbatim}
AlphaRepExpand[{k},
  {k^2 - m^2, k^2 - 2*p*k, k^2 + 2*p*k,
              k^2 - 2*q*k, k^2 + 2*q*k},
  {p^2 -> 0, q^2 -> 0, p*q -> Q^2/2},
  {Q -> 1, m^2 -> x}]
\end{verbatim}
The output states the following set of three regions:
\begin{verbatim}
{{0, 0, 0, 0, 0}, {0, 0, 0, 1, 1},
 {0, 1, 1, 0, 0}}
\end{verbatim}
As before, the regions are specified by the scaling of the Feynman parameters
in terms of powers of the small parameter~$m^2$. For example, for the
second region we have
$x_1 \sim x_2 \sim x_3 \sim(m^2)^0$, $x_4 \sim x_5 \sim(m^2)^1$.
The first region is hard; its contribution starts with $(m^2)^0$.
The second and third regions start with order $(m^2)^{-1-\ep}$. They
correspond to the two collinear regions stated for the momentum-space
expansion above. But \texttt{asy.m} does not find anything corresponding to
the Glauber region; in particular, none of the regions found by
\texttt{asy.m} provides the leading $(m^2)^{-2-\ep}$ contribution.

We notice that, as in the previous section about potential contributions,
the polynomial in the square brackets of~(\ref{Fp_integral}) has terms of
different sign. The missing Glauber contribution stems from the parameter
region where either $(x_2-x_3) \sim (m^2)^1$ or $(x_4-x_5) \sim (m^2)^1$. So let us
decompose the parametric integral into four parts corresponding to the
domains where the two factors $(x_2-x_3)$ and $(x_4-x_5)$ are either
positive or negative and then introduce new variables in such a way that
this product takes the form $\pm x'_2 x'_4$. For example, in the domain
$x_2\leq x_3$, $x_5\leq x_4$ we change the variables by
$x_2=x'_3/2,\;x_3=x'_2+x'_3/2$ and by $x_4=x'_4+x'_5/2,\;x_5=x'_5/2$,
similarly to our example in the previous section.
However, in the threshold expansion the cancelling terms appeared in
squared form such that a transformation between one pair of variables was
sufficient. Here two separate factors involve cancellations, which requires
a twofold change of variables.

Removing the primes from the variables~$x_i$, the parametric integral reads
\begin{equation}
  F(Q^2,m^2)=2(I_+ + I_-)
\end{equation}
with
\begin{align}
  I_{\pm} &=-i\pi^{d/2} \, \frac{\Gm(3+\ep)}{4}
  \idotsint \dd x_1\cdots\dd x_5
  \nn \\*  &\times
  \frac{\delta\left(x_1-1\right) \; (x_1+x_2+x_3+x_4+x_5)^{1+2\ep}
  }{\left[x_1(x_1+x_2+x_3+x_4+x_5) m^2
  \pm x_2 x_4Q^2 -i 0\right]^{3+\ep}
   }\;,
  \label{Fp_integral_A}
\end{align}
where we have chosen the argument of the delta function as $x_1-1$, so that
we may also write
\begin{align}
  I_{\pm} &=-i\pi^{d/2} \, \frac{\Gm(3+\ep)}{4}
  \int_0^\infty \cdots \int_0^\infty
  \dd x_2\cdots\dd x_5
  \nn \\*  &\times
  \frac{(1+x_2+x_3+x_4+x_5)^{1+2\ep}
  }{\left[(1+x_2+x_3+x_4+x_5) m^2
  \pm x_2 x_4Q^2 -i 0\right]^{3+\ep}
   }\;.
  \label{Fp_integral_B}
\end{align}
It is sufficient to consider the expansion of $I_+$ and obtain a result for
$I_-$ by analytically continuing $Q^2 \to -Q^2-i0$, taking into account
that the dependence on $Q^2$ is power-like.

Now we can apply {\tt asy2.m} to the integral~$I_+$ using either the command
\begin{verbatim}
WilsonExpand[x1*(x1 + x2 + x3 + x4 + x5)*m^2
    + x2*x4*Q^2,
  x1 + x2 + x3 + x4 + x5,
  {x1, x2, x3, x4, x5},
  {Q^2 -> 1, m^2 -> x}, Delta -> True]
\end{verbatim}
for integrals~(\ref{Fp_integral_A}) restricted by a delta function (see
Section~\ref{sec:syntax}), yielding the output:
\begin{verbatim}
{{0, 0, 0, 0, 0}, {0, 1, 0, 0, 0},
 {0, 0, 0, 1, 0}}
\end{verbatim}
Or we use the command
\begin{verbatim}
WilsonExpand[(1 + x2 + x3 + x4 + x5)*m^2
    + x2*x4*Q^2,
  1 + x2 + x3 + x4 + x5,
  {x2, x3, x4, x5},
  {Q^2 -> 1, m^2 -> x}]
\end{verbatim}
for integrals~(\ref{Fp_integral_B}) over variables from $0$ to $\infty$
without any restriction, and obtain the output:
\begin{verbatim}
{{0, 0, 0, 0}, {1, 0, 0, 0}, {0, 0, 1, 0}}
\end{verbatim}
The two results are equivalent, as $x_1 \sim (m^2)^0$ is implied in the
second case. So we obtain again three regions. We will see in a moment that
this list of regions is indeed correct and complete.

But first, let us emphasize that the new preresolution algorithm in
\texttt{asy2.m} is capable of performing the transformation of the integral
from (\ref{Fp_integral}) to~(\ref{Fp_integral_A}) automatically:
\begin{verbatim}
AlphaRepExpand[{k},
  {k^2 - m^2, k^2 - 2*p*k, k^2 + 2*p*k,
              k^2 - 2*q*k, k^2 + 2*q*k},
  {p^2 -> 0, q^2 -> 0, p*q -> Q^2/2},
  {Q -> 1, m^2 -> x},
  PreResolve -> True]
\end{verbatim}
The output of this command lists four different variable transformations
according to the twofold decomposition described above. For each of the
integrals over new parameters, the regions $\{0,0,0,0,0\}$, $\{0,1,0,0,0\}$
and $\{0,0,0,1,0\}$ are found, up to permutations in the order of the
parameters from different changes of variables.

The evaluation of the contributions to each region, as found by
\texttt{WilsonExpand[]} or \texttt{AlphaRepExpand[]} (including the
\texttt{PreResolve} option), is straightforward. The first region is the
hard one. The contributions of the second and third regions are not
individually regularized by dimensional regularization, as it often happens
for Sudakov-type limits. We use an auxiliary analytic regularization by
introducing additional powers $x_2^{\dl_2} x_3^{\dl_3} x_4^{\dl_4}
x_5^{\dl_5}$ of the new variables into the integrand
of~(\ref{Fp_integral_B}), taking the limit $\dl_2,\dl_3,\dl_4,\dl_5 \to 0$
in the end. The leading-order (LO) contribution of the second and third
regions to the integral $F(Q^2,m^2)$ reads
\begin{align}
  -i\pi^{d/2} \, \frac{i\pi \Gamma(\ep)}{2Q^2(m^2)^{2+\ep}}\;.
  \label{Fp_LO}
\end{align}
This agrees with the leading contribution of the Glauber region in the
momentum-space expansion.

We have found the leading Glauber contribution of order
$(m^2)^{-2-\ep}$. But we seem to have lost the two collinear regions with
the scalings $\{0,0,0,1,1\}$ and $\{0,1,1,0,0\}$ found before the change of
variables. In fact, we can evaluate the contributions from these two
regions by expanding the integral~(\ref{Fp_integral_B}). The resulting
integrals are scaleless and regularized by the parameters $\dl_3, \dl_5$,
so they vanish, and \texttt{asy2.m} is right in omitting these two regions.

We are also able to solve the integral~(\ref{Fp_integral_B}) including the
auxiliary analytic regularization factor $x_2^{\dl_2} x_3^{\dl_3}
x_4^{\dl_4} x_5^{\dl_5}$ in terms of a onefold Mellin--Barnes
representation:
\begin{align}
  I_{\pm} &=-i\pi^{d/2} \, \frac{\Gm(1+\dl_3) \Gm(1+\dl_5)}{4}
  \nn \\ &\times
  \frac{1}{2\pi i} \int \dd z \, (m^2)^z (\pm Q^2-i0)^{-3-\ep-z}
  \nn \\  &\times
  \Gm(-z) \Gm(-2-\ep+\dl_2-z) \Gm(-2-\ep+\dl_4-z)
  \nn \\  &\times
  \frac{\Gm(1-\dl_2-\dl_3-\dl_4-\dl_5+z) \Gm(3+\ep+z)}{\Gm(-1-2\ep-z)}
  \;.
  \label{Fp_integral_MB}
\end{align}
The asymptotic expansion of~$I_\pm$ in the limit $m^2/Q^2 \to 0$ is
obtained by taking the residues of the poles of the functions
$\Gm(\ldots-z)$. The poles of $\Gm(-z)$ correspond to the hard region,
while the poles of the two functions $\Gm(-2-\ep+\dl_{2,4}-z)$ provide the
contributions of the second and third regions. So \texttt{asy2.m} has found
all contributing regions.

In the Mellin--Barnes integral~(\ref{Fp_integral_MB}) we can safely take
the limit $\dl_2,\dl_3,\dl_4,\dl_5 \to 0$, add up $I_+$ and $I_-$, and
arrive at the Mellin--Barnes representation
\begin{align}
  &F(Q^2,m^2) = i\pi^{d/2} \, \frac{i}{2} \, \frac{1}{2\pi i}
  \int \dd z \, (m^2)^z (Q^2)^{-3-\ep-z} e^{i\pi(\ep+z)/2}
\nn \\*  &\;\;\times
  \Gm(-z) \Gm(-2-\ep-z) \Gm\left(\tfrac{-1-\ep-z}{2}\right) \,
  \frac{\Gm(1+z) \Gm\left(\frac{3+\ep+z}{2}\right)}{\Gm(-1-2\ep-z)}
  \;.
  \label{Fp_integral_MB2}
\end{align}
The LO contribution to $F(Q^2,m^2)$ is obtained from the residue of the
single pole at $z=-2-\ep$, in agreement with~(\ref{Fp_LO}).
The next-to-leading-order (NLO) contribution stems from the residue of the double
pole at $z=-1-\ep$ and reads
\begin{align}
  i\pi^{d/2} \, \frac{ \Gamma (1+\ep)}{(Q^2)^2(m^2)^{1+\ep}}
  \biggl( & i \frac{\pi}{2}
  +2 \psi(-\ep)- \psi(1+\ep)+ \gm_{\rm E}
\nn \\* &
  -\ln\frac{Q^2}{m^2} -1
  \biggr)\;.
  \label{Fp_NLO}
\end{align}
This agrees with the NLO contributions of the second and third regions.

At next-to-next-to-leading order (NNLO) there is a contribution from the
residue of the single pole at $z=-\ep$ which reads
\begin{equation}
  -i\pi^{d/2} \, \frac{i\pi \Gamma (2+\ep)}{4(Q^2)^3(m^2)^{\ep}}
  \label{Fp_NNLO_g}
\end{equation}
and agrees with the NNLO contributions of the second and third regions. The
second NNLO contribution comes from the residue of the single pole at
$z=0$. It is given by
\begin{equation}
  -i\pi^{d/2} \,
  \frac{i \, e^{i\pi\ep/2} \, \Gm(-2-\ep) \Gm\left(\frac{1+\ep}{2}\right)
      \Gm\left(\frac{1-\ep}{2}\right)}{
    2 (Q^2)^{3+\ep} \, \Gm(-1-2\ep)}
  \label{Fp_NNLO_h}
\end{equation}
and agrees with the LO contribution of the hard region.
So indeed all contributions to the five-point integral up to NNLO are
correctly reproduced by the contributions of the three regions found by
\texttt{asy2.m} after the decomposition of the integral and the change of
variables.

When revealing Glauber regions for a general diagram, the preresolution
algorithm of \texttt{asy2.m} tries to eliminate monomials with opposite
sign in the polynomial~$\Fc$ by automatically separating the
integration into domains and performing changes of variables. If the option
\texttt{PreResolve} is enabled for \texttt{AlphaRepExpand[]}, the code
warns the user if the elimination of monomials with opposite sign has not
been successful, such that possibly not all regions are revealed. This is
the case if some monomials of opposite sign remain in the
polynomial~$\Fc$ after tries to eliminate them, or if symbols with
unknown signs are present in the polynomial. We are therefore convinced
that \texttt{AlphaRepExpand[]}, with the preresolution enabled, either
reveals all relevant regions or issues a warning.

As for the threshold expansion of the previous section, \texttt{asy2.m} can
also treat more complicated kinematical situations, e.g. the five-point
integral depicted in Fig.~\ref{fig:Glauber} with $p_1 \ne p_2$ and $q_1 \ne
q_2$ (retaining $p_1 \parallel p_2$ and $q_1 \parallel q_2$ such that $p_i
\cdot p_j = q_i \cdot q_j = 0$ and $2p_i \cdot q_j = P_i Q_j$ with
$P_i,Q_j>0$):
\begin{verbatim}
AlphaRepExpand[{k},
  {k^2 - m^2, k^2 - 2*p1*k, k^2 + 2*p2*k,
              k^2 - 2*q2*k, k^2 + 2*q1*k},
  {p1^2 -> 0, p2^2 -> 0, p1*p2 -> 0,
    q1^2 -> 0, q2^2 -> 0, q1*q2 -> 0,
    p1*q1 -> P1*Q1/2, p1*q2 -> P1*Q2/2,
    p2*q1 -> P2*Q1/2, p2*q2 -> P2*Q2/2},
  {m^2 -> x, P1 -> 1, P2 -> 3,
             Q1 -> 2, Q2 -> 3/2},
  PreResolve -> True]
\end{verbatim}
All kinematic invariants are replaced by rational numbers in order to
enable the preresolution algorithm to work. The code correctly decomposes
the integral into four pieces and finds the three regions for each of them.

Let us finally discuss the reason why, besides the hard region which is always
present, the expansion in loop-momentum space requires two collinear
regions and one Glauber region, whereas the expansion of the parametric
integrals~(\ref{Fp_integral_B}) has two regions providing the leading
Glauber contribution and no further collinear regions.
In fact, the momentum-space expansion is also valid for loop
integrals~(\ref{5pt_integral}) where each propagator is raised to an
arbitrary, even non-integer power. For the decomposition of the parametric
integral and the change of variables, however, we have assumed the specific
form~(\ref{5pt_integral}) with each propagator present exactly once. In
this case, the loop integrand can be expanded into partial fractions as follows:
\begin{align}
  \lefteqn{\frac{1}{(k^2-m^2)(k^2-2p\cdot k) (k^2+2p\cdot k)
    (k^2-2q\cdot k) (k^2+2q \cdot k)}}
  \;
\nn \\ &
  = \frac{1}{4 (m^2)^2}
    \left( \frac{1}{k^2-m^2} - \frac{1}{k^2} - \frac{m^2}{(k^2)^2} \right)
\nn \\* &\times
    \left( \frac{1}{k^2-2p\cdot k} + \frac{1}{k^2+2p\cdot k} \right)
    \left( \frac{1}{k^2-2q\cdot k} + \frac{1}{k^2+2q\cdot k} \right)
  .
\end{align}
Expanding this product of terms, one obtains twelve three-point integrals,
which are well known. Because they only depend on $Q^2 = 2p\cdot q$, we
recognize from the last two factors the structure $F(Q^2,m^2) = 2(I_+ +
I_-)$, where $I_+$ and $I_-$ are related by $Q^2 \to -Q^2$ as before. The
three-point integrals with the massless propagators $1/k^2$ or $1/(k^2)^2$
only have a hard region. The massive three-point integral with propagator
$1/(k^2-m^2)$ is known to possess a hard and two collinear regions. Its LO
and NLO hard contributions are cancelled by the massless three-point
integrals, such that the uncancelled hard contributions start with
$(m^2)^2/(m^2)^2 = (m^2)^0$, as for the five-point integral. The LO
collinear contributions of the three-point integrals, enhanced by the
$1/(m^2)^2$ pre\-factor, scale as $(m^2)^{-2-\ep}$.

So, in the special case when all propagator powers are equal to~1, the
five-point integral reduces to a linear combination of three-point
integrals revealing the same structure of regions as found from the
expansion of the parametric integral~(\ref{Fp_integral_B}).
This picture changes when generic propagator powers are introduced as
analytic regulators, which is done in the next section.

\section{Disentangling regions via propagator powers}
\label{sec:Glauber_l}

In the previous section we have seen different patterns of regions
arising when expanding either in loop-momentum space or in parametric
space. The individual contributions can be disentangled more easily when
the dependence on the propagator powers is retained. Instead
of~(\ref{5pt_integral}), let us consider the integral
\begin{align}
  &F(Q^2,m^2)=\int \frac{\dd^d k}{(k^2-m^2)^{1+\lm_1}
    (k^2-2p\cdot k)^{1+\lm_2}}
\nn \\* &\quad\times
  \frac{1}{(k^2+2p\cdot k)^{1+\lm_3}
    (k^2-2q\cdot k)^{1+\lm_4} (k^2+2q\cdot k)^{1+\lm_5}}\;,
  \label{5pt_integral_l}
\end{align}
where the analytic regularization parameters~$\lm_i$ make the propagator
powers different from the previous case. The asymptotic expansion in
loop-momentum space yields contributions from the four regions listed in
the beginning of Section~\ref{sec:Glauber}. The LO hard contribution still
scales as $(m^2)^0$, but now the LO 1-collinear contribution scales as
$(m^2)^{-1-\ep-\lm_1-\lm_2-\lm_3}$, the LO 2-collinear contribution as
$(m^2)^{-1-\ep-\lm_1-\lm_4-\lm_5}$ and the LO Glauber contribution as
$(m^2)^{-2-\ep-\lm_1-\lm_2-\lm_3-\lm_4-\lm_5}$. So all regions are
characterized by a distinct scaling. When performing the expansion in
parametric space for generic~$\lm_i$, we are able to disentangle the
contributions from each region and match them with the regions obtained in
loop-momentum space.

For generic~$\lm_1,\ldots,\lm_5$, the parametric integral corresponding
to~(\ref{Fp_integral}) reads
\begin{align}
  \lefteqn{
  F(Q^2,m^2)=-i\pi^{d/2} \, e^{-i\pi(\lm_1+\ldots+\lm_5)}
  } \quad
\nn \\ &\times
  \frac{\Gm(3+\ep+\lm_1+\ldots+\lm_5)}{\Gm(1+\lm_1) \cdots \Gm(1+\lm_5)}
  \idotsint \dd x_1\cdots\dd x_5
\nn \\ &\times
  \delta\biggl(\sum_i x_i-1\biggr) \;
  x_1^{\lm_1} \cdots x_5^{\lm_5} \;
  (x_1+\ldots+x_5)^{1+2\ep+\lm_1+\ldots+\lm_5}
\nn \\ &\times
  \bigl[x_1(x_1+\ldots+x_5) m^2
\nn \\* &\qquad
    +(x_2-x_3)(x_4-x_5)Q^2-i 0\bigr]^{-(3+\ep+\lm_1+\ldots+\lm_5)}
  \;.
  \label{Fp_integral_l}
\end{align}
The change of variables performed in Section~\ref{sec:Glauber}, e.g.\ $x_2
= x'_2 + x'_3/2$, $x_3 = x'_3/2$ for $x_2 \ge x_3$, is complicated by the
presence of the factors $x_2^{\lm_2} x_3^{\lm_3} x_4^{\lm_4} x_5^{\lm_5}$,
where parts of the monomials will change into polynomials. If we want to
keep the simple form $F(Q^2,m^2) = 2(I_+ + I_-)$ from the previous section,
then we have to require $\lm_3 = \lm_2$ and $\lm_5=\lm_4$. Under this
restriction of the parameters~$\lm_i$, the parametric integrals can be
written as
\begin{align}
  I_\pm&=-i\pi^{d/2} \,
    \frac{e^{-i\pi(\lm_1+2\lm_2+2\lm_4)}\, \Gm(3+\ep+\lm_1+2\lm_2+2\lm_4)}{
      4^{1+\lm_2+\lm_4} \, \Gm(1+\lm_1) \Gm^2(1+\lm_2) \Gm^2(1+\lm_4)}
\nn \\ &\times
  \idotsint\dd x_1\cdots\dd x_5 \; \delta\left(x_1-1\right) \;
  x_1^{\lm_1} (2x_2+x_3)^{\lm_2} x_3^{\lm_2}
\nn \\ &\times
  \frac{(2x_4+x_5)^{\lm_4} x_5^{\lm_4} \;
  (x_1+\ldots+x_5)^{1+2\ep+\lm_1+2\lm_2+2\lm_4}
  }{\bigl[x_1(x_1+\ldots+x_5) m^2
  \pm x_2 x_4 Q^2-i 0\bigr]^{3+\ep+\lm_1+2\lm_2+2\lm_4}
   }
  \;.
  \label{Fp_integral_l_A}
\end{align}
In order to find the regions for the asymptotic expansion
of~(\ref{Fp_integral_l_A}), we have to provide the additional polynomial
factors $(2x_2+x_3)$ and $(2x_4+x_5)$ to \texttt{WilsonExpand[]}. We can do
this by multiplying the new polynomials to the second argument of the
command:
\begin{verbatim}
WilsonExpand[x1*(x1 + x2 + x3 + x4 + x5)*m^2
    + x2*x4*Q^2,
  (x1 + x2 + x3 + x4 + x5) * (2*x2 + x3)
    * (2*x4 + x5),
  {x1, x2, x3, x4, x5},
  {Q^2 -> 1, m^2 -> x}, Delta -> True]
\end{verbatim}
The output of this command reads:
\begin{verbatim}
{{0, 0, 0, 1, 0}, {0, 0, 0, 1, 1},
 {0, 1, 0, 0, 0}, {0, 1, 1, 0, 0},
 {0, 0, 0, 0, 0}}
\end{verbatim}
In addition to the regions present in the analysis of
Section~\ref{sec:Glauber}, we retrieve the two collinear regions with
scalings $\{0,0,0,1,1\}$ and $\{0,1,1,0,0\}$.
The updated code \texttt{asy2.m} is capable of taking into account generic
propagator powers automatically when told through the additional option
\texttt{GenericPowers}:
\begin{verbatim}
AlphaRepExpand[{k},
  {k^2 - m^2, k^2 - 2*p*k, k^2 + 2*p*k,
              k^2 - 2*q*k, k^2 + 2*q*k},
  {p^2 -> 0, q^2 -> 0, p*q -> Q^2/2},
  {Q -> 1, m^2 -> x},
  PreResolve -> True, GenericPowers -> True]
\end{verbatim}
When the option \texttt{GenericPowers} is enabled, the
polynomial~$\Uc$ obtained from the loop integral is multiplied by
the product of all Feynman parameters, $x_1 x_2 \cdots$, before the changes
of variables are performed. Some of these additional factors then turn into
polynomials through replacements in the preresolution algorithm, while
others remain monomials and are therefore irrelevant for the analysis of
\texttt{asy2.m}. The call of \texttt{AlphaRepExpand[]} stated above
correctly yields all five regions for each of the variable transformations.

For the evaluation of the Glauber contributions with scalings
$\{0,0,0,1,0\}$ and $\{0,1,0,0,0\}$, an additional analytic regularization
is needed, and we choose to multiply the integrand
of~(\ref{Fp_integral_l_A}) by $x_2^{\dl_2} x_4^{\dl_4}$. (The parameters
$\dl_3,\dl_5$ from Section~\ref{sec:Glauber} are not needed here due to the
presence of $\lm_2,\lm_4$.) The two Glauber contributions are individually
singular in the limit $\dl_2,\dl_4 \to 0$, but this singularity cancels in
the sum of the two contributions.

The leading contribution to the integral $F(Q^2,m^2)$ originates from the
sum of the LO Glauber contributions. It reads
\begin{align}
  &-i\pi^{d/2} \,
  \frac{i \, e^{-i\pi(\lm_1+2\lm_2+2\lm_4)} \,
      \Gm\left(\tfrac{1}{2}+\lm_2\right) \Gm\left(\tfrac{1}{2}+\lm_4\right)}{
    2 Q^2 \, (m^2)^{2+\ep+\lm_1+2\lm_2+2\lm_4}}
\nn \\* &\quad\times
  \frac{\Gm(2+\ep+\lm_1+2\lm_2+2\lm_4) \Gm(-1-\ep-2\lm_2-2\lm_4)}{
    \Gm(1+\lm_1) \Gm(1+\lm_2) \Gm(1+\lm_4) \Gm(1-\ep)}
  \;.
  \label{Fp_l_LO}
\end{align}
This agrees with the LO contribution from the one Glauber region in the
momentum-space expansion of~(\ref{5pt_integral_l}) for $\lm_3 = \lm_2$ and
$\lm_5=\lm_4$. Here, in the expansion of the parametric integrals, we have
two regions producing the Glauber contribution. This is possible because
the contributions from both regions have the same scaling (for
$\dl_2=\dl_4=0$), starting with $(m^2)^{-2-\ep-\lm_1-2\lm_2-2\lm_4}$.
For $\lm_i=0$, the result~(\ref{Fp_l_LO}) reproduces~(\ref{Fp_LO}).

Among the NLO contributions to $F(Q^2,m^2)$, we expect NLO Glauber
contributions and LO collinear contributions. However, the NLO Glauber
contributions vanish exactly for general $\lm_1$, $\lm_2=\lm_3$ and
$\lm_4=\lm_5$, due to non-trivial cancellations between the pieces which
contribute to the NLO expansion of~(\ref{Fp_integral_l_A}) for either of
the scalings $\{0,0,0,1,0\}$ and $\{0,1,0,0,0\}$. The same happens in
loop-momentum space, where the NLO Glauber contribution is proportional to
$(\lm_3-\lm_2)(\lm_5-\lm_4)$, thus vanishing in the case considered here.
So the NLO contribution to $F(Q^2,m^2)$ is made up entirely from the LO
collinear contributions. The 1-collinear region provides
\begin{align}
  &-i\pi^{d/2} \,
  \frac{e^{-i\pi(\lm_1+2\lm_2+2\lm_4)} \, e^{i\pi\lm_4} \,
      \Gm(\lm_2-\lm_4) \Gm(1-2\lm_4)}{
    2 (Q^2)^{2+2\lm_4} \, (m^2)^{1+\ep+\lm_1+2\lm_2} \,
      \Gm(1+\lm_1) \Gm(1+\lm_2)}
\nn \\* &\quad\times
  \frac{\Gm(1+\ep+\lm_1+2\lm_2) \Gm(-\ep-2\lm_2)}{\Gm(1-\lm_4) \Gm(-\ep-2\lm_4)} \;
  \frac{1}{1+2\lm_4}
  \;,
  \label{Fp_l_NLO_1c}
\end{align}
in agreement with the momentum-space expansion. The 2-collinear
contribution is obtained from this by exchanging $\lm_2 \leftrightarrow
\lm_4$. Adding the two collinear contributions together and performing the
limit $\lm_1,\lm_2,\lm_4 \to 0$, the result~(\ref{Fp_NLO}) is reproduced.

Considering finally the NNLO contributions to $F(Q^2,m^2)$, we expect NNLO
Glauber contributions, NLO collinear contributions and a LO hard
contribution. But here the NLO collinear contributions vanish exactly for
$\lm_3 = \lm_2$ and $\lm_5=\lm_4$, both in loop-momentum space and when
expanding the parametric integrals. So we are left with the NNLO Glauber
contributions yielding
\begin{align}
  &-i\pi^{d/2} \,
  \frac{i \, e^{-i\pi(\lm_1+2\lm_2+2\lm_4)} \,
      \Gm\left(\lm_2-\tfrac{1}{2}\right) \Gm\left(\lm_4-\tfrac{1}{2}\right)}{
    16 (Q^2)^3 \, (m^2)^{\ep+\lm_1+2\lm_2+2\lm_4}}
\nn \\* &\quad\times
  \frac{\Gm(\ep+\lm_1+2\lm_2+2\lm_4) \Gm(1-\ep-2\lm_2-2\lm_4)}{
    \Gm(1+\lm_1) \Gm(1+\lm_2) \Gm(1+\lm_4) \Gm(-1-\ep)}
  \label{Fp_l_NNLO_g}
\end{align}
and the LO hard contribution,
\begin{align}
  &i\pi^{d/2} \,
  \frac{i \, e^{-i\pi(\lm_1+2\lm_2+2\lm_4)} \,
      e^{i\pi(\ep+\lm_1+2\lm_2+2\lm_4)/2}}{
    2\sqrt{\pi} \, (2Q^2)^{3+\ep+\lm_1+2\lm_2+2\lm_4} \,
      \Gm(1+\lm_2) \Gm(1+\lm_4)}
\nn \\ &\quad\times
  \frac{\Gm\left(\tfrac{-1-\ep-\lm_1-2\lm_2}{2}\right)
      \Gm\left(\tfrac{-1-\ep-\lm_1-2\lm_4}{2}\right)}{
    \Gm(-1-2\ep-\lm_1-2\lm_2-2\lm_4)}
\nn \\ &\quad\times
  \Gm\left(\tfrac{-2-\ep-\lm_1-2\lm_2-2\lm_4}{2}\right)
  \Gm\left(\tfrac{3+\ep+\lm_1+2\lm_2+2\lm_4}{2}\right)
  \,,
  \label{Fp_l_NNLO_h}
\end{align}
both consistent between the expansions of the loop integral and of the
parametric integrals. Setting all $\lm_i = 0$ in (\ref{Fp_l_NNLO_g})
and~(\ref{Fp_l_NNLO_h}), we recover the results from (\ref{Fp_NNLO_g})
and (\ref{Fp_NNLO_h}), respectively.

We may also evaluate the parametric integral~(\ref{Fp_integral_l_A}) in
terms of a onefold Mellin--Barnes representation:
\begin{align}
  I_{\pm} &= -i\pi^{d/2} \,
    \frac{e^{-i\pi(\lm_1+2\lm_2+2\lm_4)}}{
      4\pi \, \Gm(1+\lm_1) \Gm(1+\lm_2) \Gm(1+\lm_4)}
\nn \\  &\times
    \frac{1}{2\pi i} \int \dd z \, (m^2)^z
    (\pm 4 Q^2-i0)^{-3-\ep-\lm_1-2\lm_2-2\lm_4-z}
\nn \\  &\times
    \frac{\Gm(1+\lm_1+z) \Gm(3+\ep+\lm_1+2\lm_2+2\lm_4+z)}{
      \Gm(-1-2\ep-\lm_1-2\lm_2-2\lm_4-z)}
\nn \\  &\times
    \Gm(-z)
    \Gm\left(\tfrac{-1-\ep-\lm_1-2\lm_2-z}{2}\right)
    \Gm\left(\tfrac{-1-\ep-\lm_1-2\lm_4-z}{2}\right)
\nn \\  &\times
    \Gm^2\left(\tfrac{-2-\ep-\lm_1-2\lm_2-2\lm_4-z}{2}\right)
  \;.
  \label{Fp_integral_l_MB}
\end{align}
The relevant regions can easily be determined from the gamma functions in
the last two lines. In particular, the squared gamma function
indicates that the expansion of the parametric integrals~$I_\pm$ requires
\emph{two} regions for the Glauber contribution, both scaling as
$(m^2)^{-2-\ep-\lm_1-2\lm_2-2\lm_4}$ at leading order.
When combining $F(Q^2,m^2) = 2(I_+ + I_-)$, one of these gamma functions is
cancelled, and we obtain
\begin{align}
  \lefteqn{
    F(Q^2,m^2) = i\pi^{d/2} \,
    \frac{i \, e^{-i\pi(\lm_1+2\lm_2+2\lm_4)}
        e^{i\pi(\ep+\lm_1+2\lm_2+2\lm_4)/2}}{
      2\sqrt{\pi} \, \Gm(1+\lm_1) \Gm(1+\lm_2) \Gm(1+\lm_4)}
  } \quad
\nn \\  &\times
    \frac{1}{2\pi i} \int \dd z \, (m^2)^z
    (2 Q^2)^{-3-\ep-\lm_1-2\lm_2-2\lm_4-z} \,
    e^{i\pi z/2}
\nn \\  &\times
    \frac{\Gm(1+\lm_1+z) \Gm\left(\tfrac{3+\ep+\lm_1+2\lm_2+2\lm_4+z}{2}\right)}{
      \Gm(-1-2\ep-\lm_1-2\lm_2-2\lm_4-z)}
\nn \\  &\times
    \Gm(-z)
    \Gm\left(\tfrac{-1-\ep-\lm_1-2\lm_2-z}{2}\right)
    \Gm\left(\tfrac{-1-\ep-\lm_1-2\lm_4-z}{2}\right)
\nn \\  &\times
    \Gm\left(\tfrac{-2-\ep-\lm_1-2\lm_2-2\lm_4-z}{2}\right)
  \;.
  \label{Fp_integral_l_MB2}
\end{align}
{}From this representation the contributions to the asymptotic expansion in
the limit $m^2/Q^2 \to 0$ can be extracted: The hard contributions stem
from the residues of the poles at $z=n$, the 1-~and 2-collinear
contributions from $z=-1+2n-\ep-\lm_1-2\lm_2$ and
$z=-1+2n-\ep-\lm_1-2\lm_4$, respectively, the Glauber contributions from
$z=-2+2n-\ep-\lm_1-2\lm_2-2\lm_4$ (with $n=0,1,2,\ldots$). All LO, NLO and
NNLO contributions reported in (\ref{Fp_l_LO})--(\ref{Fp_l_NNLO_h}) are
confirmed by the corresponding residue contributions
from~(\ref{Fp_integral_l_MB2}). In particular, the structure of the poles
in~(\ref{Fp_integral_l_MB2}) clearly shows that the Glauber region does not
contribute to the NLO result, and that the collinear contributions are
absent at NNLO, as obtained before.

The results reported in this section show that the use of generic
propagator powers helps disentangling the Glauber and collinear regions
from each other, making all regions contribute in the same way to the
asymptotic expansion in loop-momentum space and to the expansion of the
parametric integrals. Keeping the dependence of the contributions and their
scalings on the propagator powers also facilitates the identification of
regions found by \texttt{asy2.m} in parametric space for a subsequent
expansion at the level of the loop integration.

\section{Summary of \texttt{asy2.m}}
\label{sec:syntax}

The updated version of the code, \texttt{asy2.m}, can be downloaded from
the known web site \cite{asy_web}, where further installation instructions
are found.
The \texttt{Mathematica} code is loaded using \texttt{<{}<asy2.m}.

The main function \texttt{AlphaRepExpand[]} identifies all regions which
contribute to the asymptotic expansion of a given loop integral:
\begin{verbatim}
AlphaRepExpand[{k1, k2, ...},
  {(k1 + p1)^2 - m1^2, (k2 + p2)^2 - m2^2,
    ...},
  {p1^2 -> Q1, p2^2 -> Q2, p1*p2 -> Q3, ...},
  {m1^2 -> x, m2^2 -> x^2,
    Q1 -> 1, Q2 -> 3/2, ...},
  options]
\end{verbatim}
The first argument is the list of loop momenta. The second argument lists
the denominators of the propagators. The third argument contains
replacement rules for all kinematic invariants. In particular, all external
momenta appearing in the denominators must be replaced here, otherwise they
are not identified correctly as vectors. The fourth argument sets the
scaling of the parameters by replacing all symbols with powers of the
expansion parameter, labelled by the global symbol~\texttt{x}, and rational
numbers (integers or explicit fractions of integers).

The output is a list of regions, specified by the scaling (in powers of the
expansion parameter) of the Feynman parameters $x_1,x_2,\ldots$
corresponding to the propagators in the order stated in the second
argument. E.g.\ the output
\begin{verbatim}
{..., {0, 2, 1, ...}, ...}
\end{verbatim}
indicates that there is a region $\{0,2,1,\ldots\}$ specified by the
scaling $x_1 \sim x^0$, $x_2 \sim x^2$, $x_3 \sim x^1$, $\ldots$ of the
Feynman parameters, where $x$ is the small parameter of the problem.

Possible options of \texttt{AlphaRepExpand[]} are:
\begin{itemize}
\item \texttt{PreResolve -> True}: Try to eliminate cancellations between
  terms in the parametric representation by decomposing the integral and
  performing changes of variables. The output contains for each region a
  list of three entries, i.e.\ a region is e.g.\ specified by:
  \begin{verbatim}
{{x[1] -> y[1]/2, x[2] -> y[1]/2 + y[2]},
  2, {0, 1/2}}
  \end{verbatim}
  The first entry in this list is the transformation between the original
  Feynman parameters \verb@x[1]@, \verb@x[2]@, $\ldots$ and the new
  variables \verb@y[1]@, \verb@y[2]@, $\ldots$. The second entry is the
  Jacobian of the integral transformation (here ``2'', i.e.\ $\dd y_1 \dd
  y_2 = 2 \, \dd x_1 \dd x_2$). The last entry specifies the region in the
  usual form with the scalings of the new variables in powers of the small
  parameter, here $y_1 \sim x^0$, $y_2 \sim x^{1/2}$.

  When the option \texttt{PreResolve} is enabled, the code warns if it
  fails to eliminate all possible cancellations in the parametric
  representation. When no warning is issued, all regions are found. Without
  this option, however, regions occurring at such cancellations will not be
  revealed.
\item \texttt{GenericPowers -> True}: Take into account generic (in
  particular non-integer) powers of the propagators, e.g.\ when these
  powers are used as analytic regulators. Without this option, the
  preresolution algorithm triggered by the option \texttt{PreResolve} only
  finds all regions for integrals with propagators raised to positive
  integer powers.
\item \texttt{Verbose -> True}: Print verbose internal information.
\item \texttt{Scalar -> True}: Permit more complex structures of the
  denominators by specifying scalar products of momenta via the function
  \texttt{Scalar[k,p]} instead of simple products \texttt{k*p} or
  \texttt{Scalar2[k]} instead of \verb@k^2@, e.g.:
  \begin{verbatim}
AlphaRepExpand[{k},
  {Scalar[k, k] - m^2,
    Scalar2[k - q] - m^2},
  {Scalar2[q] -> qq, m^2 -> qq/4 + y},
  {qq -> 1, y -> x}, 
  PreResolve -> True, Scalar -> True]
  \end{verbatim}
\end{itemize}

For expanding more general integrals, which need not originate from Feynman
diagrams, the command \texttt{WilsonExpand[]} may be used:
\begin{verbatim}
WilsonExpand[F, U, {x1, x2, ...},
  {... -> x, ...}, options]
\end{verbatim}
Traditionally, the first two arguments are the polynomials $\Fc$ and $\Uc$,
respectively, from the alpha parametric representation~(\ref{alpha-d}) of
the Feynman integral or from the Feynman parametric representation
(\ref{alpha-d-mod}),~(\ref{integrand}). But, more generally,
\texttt{WilsonExpand[]} reveals regions for integrals over parameters
$x_1,x_2,\ldots$, integrated from 0 to~$\infty$ each, where all non-trivial
polynomials of the parameters~$x_i$ occurring in the integrand are
specified either in the first or the second argument. The third argument of
\texttt{WilsonExpand[]} is the list of integration parameters. The fourth
argument specifies the scaling of all quantities with the expansion
parameter~\texttt{x}, in the same way as in the fourth argument of
\texttt{AlphaRepExpand[]}.

The output is a list of regions, specified by the scaling of the
integration parameters in powers of the expansion parameter, exactly as
described for the output of \texttt{AlphaRepExpand[]} (without the
\texttt{PreResolve} option).

Possible options of \texttt{WilsonExpand[]} are:
\begin{itemize}
\item \texttt{Delta -> True}: Under the integral, the sum over an arbitrary
  non-empty subset of the integration parameters is restricted to~1 via a
  delta function. The specific choice of the sum in this delta function
  must be irrelevant for the integral, which is the case for the
  generalized Feynman parametric representation~(\ref{alpha-d-mod}).
\item \texttt{Verbose -> True}: Print verbose internal information.
\end{itemize}
More generally, the option \texttt{Delta} of \texttt{WilsonExpand[]} works
correctly for all integrals of the form
\begin{align}
  \int_0^\infty \cdots \int_0^\infty
  \dd x_1\cdots\dd x_N \,
  \delta\left(\sum_{i=1}^N a_i x_i - 1\right)
  f(x_1,\ldots,x_N)
  \;,
  \label{delta_int}
\end{align}
where the linear combination in the argument of the delta function has no
negative and at least one positive coefficient ($a_i \ge 0 \, \forall i$
and $\exists \, a_i>0$), and where the function~$f$ scales homogeneously
with the parameters~$x_i$ as
\begin{align}
  f(\lm x_1,\ldots,\lm x_N) = \lm^{-N} f(x_1,\ldots,x_N)
  \; \forall \lm>0
  \;,
  \label{delta_int_hom}
\end{align}
the degree of homogeneity being equal to minus the number of integration
parameters. It can be shown that for such integrals~(\ref{delta_int}) the
specific choice of the coefficients~$a_i$ is irrelevant.\footnote{%
  \label{fn:delta_scaling}%
  To see this, multiply the integrand of~(\ref{delta_int}) by~1 in the form
  $x_j \int_0^\infty \dd t \, e^{-t x_j}$, where $x_j$ is any of the
  integration parameters. Then, inside the $t$-integration, transform the
  integration variables as $x_i \to x_i/t$, $i = 1,\ldots,N$, and use the
  homogeneity relation~(\ref{delta_int_hom}) with $\lm=1/t$. Finally
  evaluate the $t$-integration first, yielding $\int_0^\infty \dd t \,
  \delta(\sum_{i=1}^N a_i x_i - t) = 1$, independent of the
  coefficients~$a_i$, as long as the linear combination is positive. The
  integral~(\ref{delta_int}) is given by $\int_0^\infty \cdots
  \int_0^\infty \dd x_1\cdots\dd x_N \, f(x_1,\ldots,x_N) \, x_j \,
  e^{-x_j}$.%
}
The integrand~(\ref{integrand}) of the Feynman parametric
representation~(\ref{alpha-d-mod}) and the integrands of all parametric
representations of Feynman integrals used in this paper (without the
additional analytic regularization factors $x_i^{\dl_i}$) fulfill the
homogeneity condition~(\ref{delta_int_hom}), so they do not depend on the
specific choice for the arguments of their delta functions.

\section{Conclusion}
\label{sec:conclusion}

We have presented an algorithm for identifying all regions which are
relevant for the asymptotic expansion of a given loop integral at the level
of its parametric representation. In contrast to the previous version
\texttt{asy.m} of the code, also potential regions and Glauber regions are
found now. The necessary decompositions and variable transformations of the
integral are automated by the updated \texttt{Mathematica} code
\texttt{asy2.m}~\cite{asy_web}. When the command \texttt{AlphaRepExpand[]}
is used with its option \texttt{PreResolve} enabled, we are convinced that
it either reveals all relevant regions or issues a warning. In particular,
regions corresponding to cancellations between large positive and negative
terms in the parametric representation of the loop integrals (such as
potential and Glauber regions) will now be found.

Let us emphasize that to prove expansion by regions
at least for some specific limit typical of Minkowski
space is a natural mathematical problem. Perhaps, this problem is not specifically
related to Feynman integrals. Let us present an example of a one-dimensional
parametric integral, without any relevance to Feynman integrals, and show that
expansion by regions works successfully. To do this, we will use
{\tt asy2.m}.

Let us consider the integral
\begin{equation}
  F(t)=\int_0^\infty(t + u + u^2)^\lm \dd u\;,
  \label{toy_example}
\end{equation}
with $\lm$ a complex parameter, in the limit $t\to 0$.
We assume that $\lm$ is in the domain Re$\lm<-1/2$ in order to have
an absolute convergence of the integral which then can be continued analytically
to the whole complex plane as an analytic function of $\lm$.
Running
\begin{verbatim}
WilsonExpand[t + u + u^2, 1, {u}, {t -> x}]
\end{verbatim}
we obtain the two regions \verb@{{1}, {0}}@.
The leading-order terms from each region
can be evaluated analytically in terms of gamma functions at general $\lm$, with
the results
\begin{equation}
  \frac{t^{\lambda +1} \Gamma (-\lambda -1)}{\Gamma (-\lambda )}
\end{equation}
and
\begin{equation}
  \frac{\Gamma (-2 \lambda -1) \Gamma (\lambda +1)}{\Gamma (-\lambda )} \;.
\end{equation}
They can be checked easily by deriving the onefold Mellin--Barnes
representation
\begin{align}
  F(t)&=\frac{1}{2\pi i}\frac{1}{\Gamma (-\lambda )}
\nn \\* &\times
  \int \Gamma (-z) \Gamma (\lambda -z+1) \Gamma (-2 \lambda +2z-1) \, t^z \, \dd z
\end{align}
and evaluating the first terms of the asymptotic expansion in the limit
$t\to 0$ by shifting the contour to the right and taking residues at the poles of
the two gamma functions in the integrand.

\begin{acknowledgements}
  This work is supported by the Deutsche For\-schungs\-ge\-mein\-schaft
  Sonder\-forschungs\-bereich/Transregio~9 ``Com\-pu\-ter\-gest\"utzte
  Theoretische Teilchenphysik''.
  The work of A.S. and V.S. is also supported by the Russian Foundation for
  Basic Research through grant 11-02-01196.
  The authors thank M.~Beneke and A.~Pak for helpful discussions.
\end{acknowledgements}

\end{document}